\documentclass[a4paper,11pt]{article}
\usepackage[dvips]{graphicx}
\usepackage{amsmath}
\usepackage{amssymb}
\usepackage{cite}
\usepackage{cancel}
\usepackage{fancybox}
\usepackage[dvips]{color}
\usepackage{ulem}
\usepackage{pifont}
\usepackage{colortbl}
\usepackage{multicol}

\usepackage{colortbl}
\usepackage{multicol}
\usepackage[all,knot]{xy}
\usepackage{longtable}

\makeatletter
 
 \@addtoreset{equation}{section}
\makeatother

\usepackage{xcolor}

\begin{document}

\begin{titlepage}

\begin{flushright}
YITP-25-45\\
\end{flushright}

\begin{center}

\vspace{0cm}
{\Large\textbf{
Family Unification in $SO(16)$ Grand Unification
}}
\vspace{1cm}

\renewcommand{\thefootnote}{\fnsymbol{footnote}}
Cheng-Wei Chiang${}^{1,2,3}$\footnote[2]{chengwei@phys.ntu.edu.tw},
Tianjun Li${}^{4,5,6}$\footnote[3]{tli@itp.ac.cn},
and Naoki Yamatsu${}^{7}$\footnote[1]{naoki.yamatsu@yukawa.kyoto-u.ac.jp}
\vspace{5mm}

\textit{
 $^1${Department of Physics and Center for Theoretical Physics,
 National Taiwan University, Taipei, Taiwan 10617, R.O.C.}\\
 $^2${Physics Division, National Center for Theoretical Sciences,
 Taipei, Taiwan 10617, R.O.C.}\\
 $^3${Kavli IPMU (WPI), UTIAS, The University of Tokyo, Kashiwa, 277-8583, Japan}\\
 $^4${CAS Key Laboratory of Theoretical Physics, Institute of
 Theoretical Physics,\\
 Chinese Academy of Sciences, Beijing 100190, P. R. China}\\
 $^5${School of Physical Sciences, University of Chinese Academy of
 Sciences,\\
 No. 19A Yuquan Road, Beijing 100049, P. R. China}\\
 $^6${School of Physics, Henan Normal University, Xinxiang 453007, P. R. China}\\
 $^7${Yukawa Institute for Theoretical Physics, Kyoto University,\\
 Kitashirakawa Oiwakecho, Sakyo-ku, Kyoto 606-8502, Japan}
}

\date{\today}

\abstract{
We propose a unified model for the three Standard Model (SM) gauge
symmetries and $SU(3)$ family symmetry based on $SO(16)$ grand unified gauge
symmetry on six-dimensional (6D) spacetime.
In this model, three chiral generations of quarks and leptons are unified into a 6D Weyl fermion in the spinor representation of $SO(16)$.
The 6D $SO(16)$ gauge anomaly is canceled by the vectorlike nature of
 the model, and the 4D gauge anomalies are canceled by introducing
 suitable 4D localized fermions at the fixed point.
The 4D gauge coupling constant of $SO(16)$ has the property of becoming weaker at high energies.
}

\end{center}
\end{titlepage}


\section{Introduction}
\label{Sec:Introduction}

The existence of three chiral generations of quarks and leptons is one
of the most mysterious facts in particle physics.
In addition, the mass hierarchy among different flavors of fermions generated by
the Higgs mechanism via their corresponding Yukawa couplings strongly
calls for a better understanding of the underlying structure of Nature.
There are many attempts to understand the origin of chiral generations
and/or their mass hierarchy by considering, e.g., the
so-called horizontal symmetry (or family symmetry) in four-dimensional
(4D) theories 
\cite{Wilczek:1978xi,Froggatt:1978nt,Yanagida:1979as,Maehara:1979kf,Inoue:1994qz,King:2001uz,Maekawa2004,Yamatsu:2007,Yamatsu:2012}, as well as employing special
geometrical structures in higher-dimensional theories
\cite{Yoshioka:1999ds,Fujimoto:2012wv}.

It is well-known that the quarks and leptons for each generation in the Standard
Model (SM) can be unified into one multiplet (or two multiplets) in
grand unified theories (GUTs) in the 4D
\cite{Georgi:1974sy,Inoue:1977qd,Fritzsch:1974nn,Ida:1980ea,Fujimoto:1981bv,Gursey:1975ki}
and higher-dimensional frameworks
\cite{Kawamura:1999nj,Kawamura:2000ir,Kawamura:2000ev,Hall:2001pg,Burdman:2002se,Lim:2007jv,Kim:2002im,Fukuyama:2008pw,Hosotani:2015hoa}.

There are some attempts to unify GUT and family groups into a larger GUT
group in 4D and higher-dimensional theories 
\cite{Ramond:1979py,Arkani-Hamed:2001vvu,Kawamura:2007cm,Kawamura:2009gr,Goto:2013jma,Fonseca:2015aoa,Albright:2016lpi,Goto:2017zsx,Reig:2017nrz,Yamatsu:2018fsg,Ekstedt:2020gaj,Maru:2024ghd}.
In particular, models based on $SU(19)$ GUT gauge symmetry are discussed
in the 4D framework \cite{Fonseca:2015aoa,Ekstedt:2020gaj} 
and the six-dimensional (6D) framework by one of the authors
\cite{Yamatsu:2018fsg}. In these models, the three chiral generations of the
SM fermions are unified into a fermion multiplet in the second-rank antisymmetric
tensor representation ${\bf 171}$ of $SU(19)$.
In such models, $SU(19)$ is broken to a maximal regular subgroup 
$SU(16)\times SU(3)\times U(1)$, and $SU(16)$ is further broken to a
maximal special subgroup $SO(10)$. The branching rule of ${\bf 171}$
under $SU(19)\supset SO(10)\times SU(3)\times U(1)$ is given by
\cite{Yamatsu:2015gut}
\begin{align}
 {\bf 171}=({\bf 16,3})(-13)\oplus {\bf (120,1)}(6)
\oplus{(\bf 1,\overline{3})}(-32),
\end{align}
where the $SO(10)$ ${\bf 120}$ representation is real.
That is, only the first term is a complex representation of $SO(10)$.
Therefore, when we introduce a fermion in ${\bf 171}$ of $SU(19)$,
only the three chiral generations of SM fermions appear 
after we take into account $SU(3)$ and $U(1)$ breaking effects.
For more information about this type of symmetry breaking, see
Refs.~\cite{Yamatsu:2015gut,Yamatsu:2017sgu,Yamatsu:2017ssg,Yamatsu:2018fsg}.

For $SO(N)$ GUT gauge symmetry, the models are inconsistent with
low-energy phenomena because additional generations or mirror particles
would appear in the 4D framework.
There have been discussions aiming at unifying the three chiral generations
of SM fermions in the framework of 5-dimensional (5D) orbifold spacetime
\cite{Reig:2017nrz}.
In Ref.~\cite{Reig:2017nrz}, unification of the SM gauge symmetry and
family symmetry based on $SO(18)$ GUT gauge symmetry is discussed in the
5D spacetime of $M^{4}\times S_1/(\mathbb{Z}_2\times \mathbb{Z}_2')$.
The $SO(18)$ gauge symmetry is broken to $SO(10)\times SO(8)$ by
the orbifold.
Besides, the confinement of $SO(5)$, which is a subgroup of $SO(8)$, is used along with an orbifold space structure to remove the mirror SM fermions.
One of the most subtle points in this model is the use of the branching
rule for the vector representation as the branching rule for the spinor
representation of $SO(8)$.
Furthermore, the $SO(18)$ gauge symmetry is used in this model for
this reason to achieve the $SO(5)$ confinement. 
If we do not need to use confinement, the $SO(16)$ gauge symmetry should
be sufficient to unify the three chiral generations of SM fermions.
This is because the branching rules of 
$SO(16)\supset SO(10)\times SU(3)\times U(1)$ for spinor representations 
${\bf 128}$ and ${\bf 128'}$ are given by
\begin{align}
&{\bf 128} =({\bf 16,3})(1)\oplus ({\bf 16,1})(-3)\oplus 
\oplus ({\bf \overline{16},\overline{3}})(-1)
\oplus ({\bf \overline{16},1})(3),
\nonumber\\
&{\bf 128'} =({\bf 16,\overline{3}})(-1)
\oplus ({\bf 16,1})(3)
\oplus ({\bf \overline{16},3})(1)
\oplus ({\bf \overline{16},1})(-3).
\label{Eq:Branching_rule-SO16_SO10-SU3-U1}
\end{align}
(For more information about Lie algebras and their subalgebras, see,
e.g., Ref.~\cite{Yamatsu:2015gut}.)

In this paper, we will propose a new model based on $SO(16)$ GUT gauge
symmetry that unifies the three chiral generations of SM fermions
into a 6D Weyl fermion in a spinor representation of
$SO(16)$ without exotic fermions on the 6D spacetime 
$M^{4}\times D^2/\mathbb{Z}_N$ $(N\geq 9)$,
where $D^2$ is a 2-dimensional (2D) disk and there is a fixed
point at the origin \cite{Li:2001dt}.
The 6D $SO(16)$ gauge anomaly is canceled by the vectorlike nature of
the model, and the 4D gauge anomalies are canceled by introducing
suitable 4D localized fermions at the fixed point.
The gauge coupling constant of $SO(16)$ is asymptotically free.

This paper is organized as follows.
In Sec.~\ref{Sec:SO16}, we introduce 
the 6D spacetime $M^4\times D^2/\mathbb{Z}_N$ and 
$SO(16)$ gauge symmetry along with its breaking.
In Sec.~\ref{Sec:Gauge}, we discuss the relations among the symmetries, gauge
bosons, and orbifolding the extra space $D^2/\mathbb{Z}_N$. 
In Sec.~\ref{Sec:Fermion}, we show how to realize the three chiral
generations of SM fermions without exotic fermions
on $M^4\times D^2/\mathbb{Z}_N$.
In Sec.~\ref{Sec:Anomaly}, 
we examine how to realize 6D and 4D gauge anomaly cancellation in the
model.
In Sec.~\ref{Sec:Symmetry-Breaking},
we demonstrate how to realize the breakdown of $SO(16)$ GUT symmetry to the SM
gauge symmetry.
In Sec.~\ref{Sec:RGE},
we analyze the renormalization group equation (RGE) for the gauge
coupling constant of $SO(16)$.
Section~\ref{Sec:Summary} summarizes and discusses our findings in this work.

\section{6D spacetime and $SO(16)$ gauge symmetry}
\label{Sec:SO16}

Let's consider the 6D spacetime $M^4\times D^2/\mathbb{Z}_N$ with Cartesian coordinates $x^\mu$ ($\mu=0,1,2,3$)  and circular coordinates $(r,\theta)$ or a complex coordinate $z=re^{i\theta}$, where $D^2$ is a 2-dimensional (2D) disk and there is a fixed point at the origin $r=0$ \cite{Li:2001dt}.
We take the radius of the disc $D^2$ as $R$.
A $\mathbb{Z}_N$ symmetry on $D^2$ is defined as
\begin{align}
 \omega=e^{i\frac{2\pi}{N}},
\end{align}
where the generator for $\mathbb{Z}_N$, $\Omega$, satisfies $\Omega^N=1$.
The metric of the 6D spacetime is given by
\begin{align}
ds^2=\eta_{\mu\nu}dx^\mu dx^\nu+ dr^2+r^2d\theta^2,
\end{align}
where $\eta_{\mu\nu}=\mbox{diag}(-1,1,1,1)$.

In the following, we discuss the gauge fields associated with the $SO(16)$ generators
$T^\alpha$ ($\alpha=1,2,...,120$), which are expressed using the Pauli
matrices $\sigma_a$ ($a=1,2,3$) and $8\times 8$ symmetric and
anti-symmetric matrices $S_{8}$ and $A_{8}$ as \cite{Huang:2004ui}:
\begin{align}
T^\alpha= 
\{\sigma_0,\sigma_1,\sigma_3\}\otimes A_8,\ \ \sigma_2\otimes S_8,
\end{align}
where $\sigma_0$ is the $2\times 2$ identity matrix.
The numbers of $S_{8}$ and $A_{8}$ matrices are $36$ and $28$, respectively.
(The total number of the generators is therefore $3\times 28+36=120$.)
Denoting the corresponding components of the gauge field as 
$A_M^{\sigma_x}$ $(x=0,1,2,3)$, we can express the gauge field $A_M$ as 
\begin{align}
A_M=
\left(
\begin{array}{cc}
A_M^{\sigma_0}+iA_M^{\sigma_3}&
A_M^{\sigma_1}+iA_M^{\sigma_2}\\
A_M^{\sigma_1}-iA_M^{\sigma_2}&
A_M^{\sigma_0}-iA_M^{\sigma_3}\\
\end{array}
\right).
\label{Eq:A_M}
\end{align}

Here we consider $SO(16)$ symmetry breaking by the $\mathbb{Z}_N$ orbifold.
First, the unbroken generators $T^a$ and the broken generators
$T^{\hat{a}}$ satisfy respectively
\begin{align}
R_\Omega T^aR_\Omega^{-1} =+T^a,\ \ \
R_\Omega T^{\hat{a}}R_\Omega^{-1} \not=T^{\hat{a}},
\end{align}
where $R_\Omega$ denotes an element in the adjoint representation of the bulk gauge group $G$ that satisfies $R_\Omega^N=I_n$, where $I_n$ is the $n\times n$ identity matrix.
For a generic bulk multiplet $\Phi$ in a representation of $G$, we have
\begin{align}
\Omega\Phi(x^\mu,z,\bar{z}) =
\Phi(x^\mu,\omega z,\bar{\omega}\bar{z})
=\eta_\Phi
\left(R_\Omega\right)^{\ell_\Phi}
\Phi(x^\mu,z,\bar{z})
\left(R_\Omega^{-1}\right)^{m_\Phi},
\end{align}
where $\bar{\omega}:=\omega^{-1}$, $\eta_\Phi\subset \mathbb{Z}_N$, and $\ell_\Omega$ and $m_\Omega$ are non-negative integers depending on the representation of $G$.
The boundary condition at $r=R$ in $D^2/\mathbb{Z}_N$ can be chosen as the Dirichlet or Neumann boundary condition. However, if the Dirichlet boundary condition is chosen, there will be no zero mode, so in the ensuing discussion, the Neumann boundary condition will be chosen to allow the existence of zero modes.
(For details, see, e.g., Refs.~\cite{Arkani-Hamed:2001vvu,Li:2001dt}.)
We will apply this discussion for the $SO(16)$ gauge field $A_\mu$ in the 4D spacetime below, noting that different choices of $R_\Omega$ will lead to different symmetry-breaking schemes.

First, we consider that $SO(16)$ is broken to $SU(8)\times U(1)_a$ by
the $\mathbb{Z}_N$ orbifold. The branching rules of 
$SO(16)\supset SU(8)\times U(1)_a$ for some representations are given by \cite{Yamatsu:2015gut}
\begin{align}
{\bf 16}&=({\bf 8})(1)\oplus({\bf \overline{8}})(-1),\nonumber\\
{\bf 120}&=({\bf 63})(0)\oplus({\bf 1})(0)
\oplus({\bf 28})(2)\oplus({\bf \overline{28}})(-2)
,\nonumber\\
{\bf 128}&=({\bf 56})(-1)\oplus({\bf \overline{56}})(1)\oplus({\bf 8})(-3)\oplus({\bf \overline{8}})(3),\nonumber\\
{\bf 128'}&=({\bf 70})(0)\oplus({\bf 28})(-2)\oplus({\bf \overline{28}})(2)\oplus({\bf 1})(4)\oplus({\bf 1})(-4).
\end{align}
This symmetry-breaking pattern is realized by taking the following matrix:
\begin{align}
R_\Omega=
\mbox{exp}\left[
i\frac{2\pi}{N}\sigma_2\otimes (-a)I_8
\right],
\end{align}
where $a=1,2,\cdots,N-1$ (mod. $N$), 
and the exponents are proportional to the $U(1)_a$ charges.
In this case, the gauge field $A_\mu$ is transformed under $\mathbb{Z}_N$
as 
\begin{align}
A_\mu=
\left(
\begin{array}{cc}
+A_\mu^{\sigma_0}+iA_\mu^{\sigma_3}&
+A_\mu^{\sigma_1}+iA_\mu^{\sigma_2}\\
+A_\mu^{\sigma_1}-iA_\mu^{\sigma_2}&
+A_\mu^{\sigma_0}-iA_\mu^{\sigma_3}\\
\end{array}
\right)
\ \to\ 
R_\Omega A_\mu R_\Omega^{-1}=
\left(
\begin{array}{cc}
+A_\mu^{\sigma_0}-iA_\mu^{\sigma_3}&
-A_\mu^{\sigma_1}+iA_\mu^{\sigma_2}\\
-A_\mu^{\sigma_1}-iA_\mu^{\sigma_2}&
+A_\mu^{\sigma_0}+iA_\mu^{\sigma_3}\\
\end{array}
\right).
\end{align}
The $\sigma_0$ and $\sigma_2$ component fields $A_\mu^{\sigma_0}$ and
$A_\mu^{\sigma_2}$ are invariant under $\mathbb{Z}_N$, while the
$\sigma_1$ and $\sigma_3$ component fields  $A_\mu^{\sigma_1}$ and
$A_\mu^{\sigma_3}$ are not invariant. Therefore, $SO(16)$ is broken to
$SU(8)\times U(1)_a$.

Next, we consider $SO(16)$ breaking to $SU(5)\times SU(3)\times U(1)_a\times U(1)_b$ by the $\mathbb{Z}_N$ orbifold. 
The branching rules of $SU(8)\supset SU(5)\times SU(3)\times U(1)_b$ for some representations
are given by \cite{Yamatsu:2015gut}
\begin{align}
{\bf 8}&=({\bf 5,1})(3)\oplus({\bf 1,3})(-5),\nonumber\\
{\bf \overline{8}}&=({\bf \overline{5},1})(-3)\oplus({\bf 1,\overline{3}})(5),\nonumber\\
{\bf 28}&=({\bf 10,1})(6)\oplus({\bf 5,3})(-2)\oplus({\bf 1,\overline{3}})(-10)
,\nonumber\\
{\bf \overline{28}}&=({\bf \overline{10},1})(-6)\oplus({\bf \overline{5},\overline{3}})(2)\oplus({\bf 1,3})(10)
,\nonumber\\
{\bf 56}&=({\bf \overline{10},1})(9)\oplus({\bf 10,3})(1)\oplus({\bf 5,\overline{3}})(-7)\oplus({\bf 1,1})(-15)
,\nonumber\\
{\bf \overline{56}}&=({\bf 10,1})(-9)\oplus({\bf \overline{10},\overline{3}})(-1)\oplus({\bf \overline{5},3})(7)\oplus({\bf 1,1})(15)
,\nonumber\\
{\bf 63}&=({\bf 24,1})(0)\oplus({\bf 1,8})(0)\oplus({\bf 1,1})(0)
\oplus({\bf 5,\overline{3}})(8)\oplus({\bf \overline{5},3})(-8)
,\nonumber\\
{\bf 70}&=({\bf 10,\overline{3}})(-4)\oplus({\bf \overline{10},3})(4)\oplus({\bf 5,1})(-12)\oplus({\bf \overline{5},1})(12).
\end{align}
We find that the following matrix realizes 
$SO(16)$ breaking to $SU(5)\times SU(3)\times U(1)_a\times U(1)_b$:
\begin{align}
R_\Omega=
\mbox{exp}\left[
i\frac{2\pi}{N}
\sigma_2\otimes
\left\{
\left(-a-3b\right)I_5
\oplus \left(-a+5b\right)I_3
\right\}
\right],
\label{Eq:R_Omega-SO16-SU5-SU3-U1-U1}
\end{align}
where $|a+3b|,|a-5b|\in \mathbb{Z}_{\not=0}$, and the exponents are
proportional to the $U(1)_a$ and $U(1)_b$ charges.

Finally, we consider $SO(16)$ breaking to $SO(10)\times SU(3)\times U(1)$ by the $\mathbb{Z}_N$ orbifold.  The branching rules of $SO(16)\supset SO(10)\times SU(3)\times U(1)$ for some representations are already given in Eq.~(\ref{Eq:Branching_rule-SO16_SO10-SU3-U1}) and by
\cite{Yamatsu:2015gut}
\begin{align}
{\bf 16}&=({\bf 10,1})(0)\oplus({\bf 1,\overline{3}})(2)
\oplus({\bf 1,3})(-2),\nonumber\\
{\bf 120}&=({\bf 45,1})(0)
\oplus({\bf 1,8})(0)\oplus({\bf 1,1})(0)
\oplus({\bf 1,3})(4)\oplus({\bf 1,\overline{3}})(-4)
\oplus({\bf 10,3})(-2)
\oplus({\bf 10,\overline{3}})(2).
\end{align}
This symmetry-breaking pattern is realized by choosing $a=-3\ell/4$ and
$b=\ell/4$ $(\ell\in \mathbb{Z}_{\not=0})$ in
Eq.~(\ref{Eq:R_Omega-SO16-SU5-SU3-U1-U1}) as 
\begin{align}
R_\Omega=
\mbox{exp}\left[
i\frac{2\pi}{N}
\sigma_2\otimes
\left\{
0\cdot 
I_5
\oplus 2\ell \cdot I_3
\right\}
\right],
\end{align}
where the exponents are proportional to the $U(1)$ charges.
This matrix breaks $SO(16)$ to $SO(10)\times SU(3)\times U(1)$.
Note that if we choose special values of $N$ and $\ell$, then $SO(16)$ is unbroken for $\ell=0$ (mod.~$N/2$), and 
$SO(16)$ is broken to $SO(10)\times SO(6)\simeq SO(10)\times SU(4)$
for $\ell=N/4$ (mod.~$N/2$), where $\ell\in \mathbb{Z}_{\not=0}$.
We will consider the $SO(16)$ breaking to $SO(10)\times SU(3)\times U(1)$ in the following sections.

\section{Gauge symmetry and gauge boson}
\label{Sec:Gauge}

In this section, we discuss the conditions that $SO(16)$ GUT gauge symmetry can be broken to $SO(10)\times SU(3)\times U(1)$.  We also find the conditions that no extra scalar particles emerge
from the extra-dimensional components of the gauge boson.

The gauge boson transforms under $\Omega$ as
\cite{Arkani-Hamed:2001vvu,Li:2001dt}
\begin{align}
&\Omega A_\mu(x^\mu,z,\bar{z})
=A_\mu(x^\mu\omega z,\bar{\omega}\bar{z})=
R_\Omega  A_\mu(x^\mu,z,\bar{z})R_\Omega^{-1},
\nonumber\\
&\Omega A_z(x^\mu,z,\bar{z})
=A_z(x^\mu,\omega z,\bar{\omega}\bar{z})=
\omega^{-1}
R_\Omega  A_z(x^\mu,z,\bar{z})R_\Omega^{-1}.
\end{align}
The phase factors of the components of the gauge field $A_M$ $(M=\mu,z)$
are summarized in Table~\ref{Tab:BC-gauge}, where
$G_{10,3,1}:=SO(10)\times SU(3)\times U(1)$.

\begin{table}[tbh]
\begin{center}
\begin{tabular}{cccc}\hline
$G_{10,3,1}$&$A_\mu$&$A_z$\\
\hline
$({\bf 45,1})(0)$
&$1$&$\omega^{-1}$\\
$({\bf 1,8})(0)$
&$1$&$\omega^{-1}$\\
$({\bf 1,1})(0)$
&$1$&$\omega^{-1}$\\
[0.5em]
$({\bf 1,3})(+4)$
&$\omega^{+4\ell}$&$\omega^{-1+4\ell}$\\
$({\bf 1,\overline{3}})(-4)$
&$\omega^{-4\ell}$&$\omega^{-1-4\ell}$\\
[0.5em]
$({\bf 10,\overline{3}})(+2)$
&$\omega^{+2\ell}$&$\omega^{-1+2\ell}$\\
$({\bf 10,{3}})(-2)$
&$\omega^{-2\ell}$&$\omega^{-1-2\ell}$\\
[0.25em]
\hline
\end{tabular}
\caption{\small
Phase factors of the components of the gauge field $A_M$ $(M=\mu,z)$, where $\ell$ is an integer.
}
\label{Tab:BC-gauge}
\end{center}
\end{table}

From Table~\ref{Tab:BC-gauge}, we find that there are the following
symmetry breaking patterns, depending on the value of $\ell$:
\begin{align}
SO(16)\to
\left\{
\begin{array}{ll}
SO(16)
&\mbox{for}\ \ell=0\ (\mbox{mod.}\ N/2), \\
SO(10)\times SU(4)
&\mbox{for}\ \ell=N/4\ (\mbox{mod.}\ N/2), \\
SO(10)\times SU(3)\times U(1)
&\mbox{for the others}.
\end{array}
\right.
\label{Eq:SO16-Symmetry-breaking}
\end{align}

We focus on the symmetry-breaking pattern 
$SO(16)\to SO(10)\times SU(3)\times U(1)$
in Eq.~(\ref{Eq:SO16-Symmetry-breaking}), where the three chiral generations
of SM fermions can be unified into one fermion multiplet.
In this case, we can take $\ell\not=0$ (mod. $N/4$).
Here, we further impose the condition that no extra zero mode appears
from the extra-dimensional components of the gauge field.
From Table~\ref{Tab:BC-gauge}, we find the following constraints for
$\ell$ and $N$ to avoid the unwanted zero modes:
\begin{align}
\pm2\ell,\
\pm4\ell,\
-1\pm2\ell,\
-1\pm4\ell
\not=0\ (\mbox{mod.}\ N).
\label{Eq:SO10-constraint-gauge}
\end{align}
Obviously, we cannot take $\ell=0$.  We next consider the $\ell=1$ case.  From Eq.~(\ref{Eq:SO10-constraint-gauge}), we find
\begin{align}
\pm2,\
\pm4,\
+1,\
-3,\
+3,\
-5
\not=0\ (\mbox{mod.}\ N).
\label{Eq:SO10-constraint-gauge-ell-1}
\end{align}
This leads to the requirement that $N\geq 6$.
For $\ell=2$, from Eq.~(\ref{Eq:SO10-constraint-gauge}),
we find
\begin{align}
\pm4,\
\pm8,\
+3,\
-5,\
+7,\
-9
\not=0\ (\mbox{mod.}\ N).
\label{Eq:SO10-constraint-gauge-ell-2}
\end{align}
This leads to $N=6$ or $N\geq 10$.
We can also consider $\ell \geq 3$ and negative integers in the same manner.

\section{Unification of the three chiral generations of SM fermions}
\label{Sec:Fermion}

We introduce 6D positive and negative Weyl fermions
$\Psi_{\bf 128}^{\alpha=1}(x,z,\bar{z})$ and 
$\Psi_{\bf 128}^{\alpha=2}(x,z,\bar{z})$ 
in a spinor representation ${\bf 128}$ of $SO(16)$.
The purpose of having the ``positive'' and ``negative'' Weyl fermions is to render a vectorlike fermion in the 6D gauge theory.
These 6D Weyl fermions can be decomposed into 4D Weyl
fermions as
\begin{align}
\Psi_{\bf 128}^{\alpha}(x,z,\bar{z})
=
\left(
\begin{array}{c}
\psi_{L{\bf 128}}^{\alpha}\\
\psi_{R{\bf 128}}^{\alpha}\\
\end{array}
\right),
\end{align}
where $\psi_{L{\bf 128}}^{\alpha}$ and $\psi_{R{\bf 128}}^{\alpha}$
stand for 4D left- and right-handed Weyl fermions,
respectively. 
The transformations of fermions under $\Omega$ are given by 
\cite{Arkani-Hamed:2001vvu,Li:2001dt}
\begin{align}
\Omega\Psi_{\bf 128}^{\alpha}(x^\mu,z,\bar{z})
=\Psi_{\bf 128}^{\alpha}(x^\mu,\omega z,\bar{\omega}\bar{z})
=\eta_{\Psi^\alpha} R_\Omega^{\rm sp}
\Psi_{\bf 128}^{\alpha}(x^\mu,z,\bar{z}),
\label{Eq:fermion-omega}
\end{align}
where
$\eta_{\Psi^1}=\omega^{m}$, $\eta_{\Psi^2}=\omega^{k}$, with
$m,k=0,1,2,...,N-1$ $(\mbox{mod.}\ N)$.
For the 6D negative Weyl fermion
$\Psi_{\bf 128}^{\alpha=2}(x,y,v)$, there are no zero modes.
The phase factors of $\Psi_{\bf 128}^{\alpha}$
$(\alpha=1,2)$ in the basis of 
$G_{10,3,1}:=SO(10)\times SU(3)\times U(1)$
are shown in Table~\ref{Tab:BC-fermion-spinor}, where 
each phase factor is given by 
$\eta_{\Psi^\alpha}R_\Omega^{\rm sp}$.
Note that the 4D right-handed fermions can be written in terms of conjugates of the 4D left-handed fermions, and the gauge symmetry imposes $\eta_{\Psi^\alpha}\eta_{\Psi^\alpha}'=\omega$, where we denote the phase factor of the left-handed fermions as $\eta_{\Psi^\alpha}$ and that of the conjugate of the left-handed fermions as $\eta_{\Psi^\alpha}'$.

\begin{table}[thb]
{
\renewcommand{\arraystretch}{1.05}
\begin{center}
\begin{tabular}{c}
$\Psi_{\bf 128}^{\alpha}$\\
\begin{tabular}{ccc}\hline
$G_{10,3,1}$&Left&Left$^c$\\\hline
$({\bf 16,3})(+1)$
&$\eta_{\Psi^\alpha}\omega^{+\ell}$&$\eta_{\Psi^\alpha}'\omega^{+\ell}$\\
$({\bf 16,1})(-3)$
&$\eta_{\Psi^\alpha}\omega^{-3\ell}$&$\eta_{\Psi^\alpha}'\omega^{-3\ell}$\\
$({\bf \overline{16},\overline{3}})(-1)$
&$\eta_{\Psi^\alpha}\omega^{-\ell}$&$\eta_{\Psi^\alpha}'\omega^{-\ell}$\\
$({\bf \overline{16},1})(+3)$
&$\eta_{\Psi^\alpha}\omega^{+3\ell}$&$\eta_{\Psi^\alpha}'\omega^{+3\ell}$\\
\hline
\end{tabular}
\end{tabular}
\caption{\small
Phase factors of the components of the fermions
$\Psi_{\bf 128}^{\alpha}$
$(\alpha=1,2)$, where $\eta_{\Psi^\alpha}\eta_{\Psi^\alpha}'=\omega$.
}
\label{Tab:BC-fermion-spinor}
\end{center}
}
\end{table}

Here we consider the case of $SO(16)$ breaking to 
$SO(10)\times SU(3)\times U(1)$, where we take $\ell\not=0$ (mod.~$N/4$).
For $\Psi_{\bf 128}^{\alpha=1}$ with $\eta_{\Psi^1}=\omega^{-\ell}$,
we find the following constraints for $\ell$ from the left table in
Table~\ref{Tab:BC-fermion-spinor-SO10}
to realize the three chiral generations of SM fermions without 
producing exotic fermions:
\begin{align}
-4\ell,\
\pm 2\ell,\
\pm2\ell+1,\
1,\
4\ell+1
\not=0\ (\mbox{mod.}\ N).
\label{Eq:SO10-constraint-fermion-1}
\end{align}
Also, we need to take $\eta_{\Psi^2}(=\omega^{k})$ not to produce zero
modes for $\Psi_{\bf 128}^{\alpha=2}$. From the right table in
Table~\ref{Tab:BC-fermion-spinor-SO10},
we find the following constraints for $k$ and $\ell$:
\begin{align}
k\pm \ell,\
k\pm 3\ell,\
-k\pm \ell+1,\
-k\pm 3\ell+1
\not=0\ (\mbox{mod.}\ N).
\label{Eq:SO10-constraint-fermion-2}
\end{align}

\begin{table}[htb]
{
\renewcommand{\arraystretch}{1.05}
\begin{center}
\begin{tabular}{cc}
$\Psi_{\bf 128}^{\alpha=1}$&
$\Psi_{\bf 128}^{\alpha=2}$\\
\begin{tabular}{ccc}\hline
$G_{10,3,1}$&Left&Left$^c$\\\hline
$({\bf 16,3})(+1)$
&$1$&$\omega^{+2\ell+1}$\\
$({\bf 16,1})(-3)$
&$\omega^{-4\ell}$&$\omega^{-2\ell+1}$\\
$({\bf \overline{16},\overline{3}})(-1)$
&$\omega^{-2\ell}$&$\omega^{+1}$\\
$({\bf \overline{16},1})(+3)$
&$\omega^{+2\ell}$&$\omega^{+4\ell+1}$\\
\hline
\end{tabular}
&
\begin{tabular}{ccc}\hline
$G_{10,3,1}$&Left&Left$^c$\\\hline
$({\bf 16,3})(+1)$
&$\omega^{+k+\ell}$&$\omega^{-k+\ell+1}$\\
$({\bf 16,1})(-3)$
&$\omega^{+k-3\ell}$&$\omega^{-k-3\ell+1}$\\
$({\bf \overline{16},\overline{3}})(-1)$
&$\omega^{+k-\ell}$&$\omega^{-k-\ell+1}$\\
$({\bf \overline{16},1})(+3)$
&$\omega^{+k+3\ell}$&$\omega^{-k+3\ell+1}$\\
\hline
\end{tabular}
\end{tabular}
\caption{\small
Phase factors of the components of the fermions
$\Psi_{\bf 128}^{\alpha}$ $(\alpha=1,2)$, where
$\eta_{\Psi^\alpha}\eta_{\Psi^\alpha}'=\omega$.
$\eta_{\Psi^1}=\omega^{-\ell}$ and $\eta_{\Psi^2}=\omega^{+k}$.
}
\label{Tab:BC-fermion-spinor-SO10}
\end{center}
}
\end{table}

We need to find a set of $\ell$, $k$ and $N$ that simultaneously satisfy 
Eqs.~(\ref{Eq:SO10-constraint-fermion-1}) and (\ref{Eq:SO10-constraint-fermion-2}).
There turns out to be an infinite number of allowed sets of $\ell$, $k$ and $N$.  Hence, 
we will only consider a few simple examples.
First, we set $\ell=1$. In this case, 
Eq.~(\ref{Eq:SO10-constraint-fermion-1}) gives
\begin{align}
-4,\
\pm 2,\
3,\
-1,\
+1,\
+5
\not=0\ (\mbox{mod.}\ N).
\end{align}
We thus need to choose $N\geq 6$.
On the other hand, Eq.~(\ref{Eq:SO10-constraint-fermion-2}) has
\begin{align}
k\pm 1,\
k\pm 3,\
-k+2,\
-k,\
-k+4,\
-k-2
\not=0\ (\mbox{mod.}\ N).
\end{align}
Taking the example of $k=-4$, we have the condition as
\begin{align}
-3,\
-5,\
-1,\
-7,\
+6,\
+4,\
+8,\
+2
\not=0\ (\mbox{mod.}\ N).
\end{align}
This leads to the requirement $N\geq 9$. 
In other words, we have now found a possible set of parameters, $(\ell,k,N)=(1,-4,9)$.

We next try $\ell=2$.  In this case, Eq.~(\ref{Eq:SO10-constraint-fermion-1}) gives
\begin{align}
-8,\
\pm 4,\
+5,\
-3,\
1,\
9
\not=0\ (\mbox{mod.}\ N).
\end{align}
This leads to $N=6,7$ or $N\geq 10$.
On the other hand, Eq.~(\ref{Eq:SO10-constraint-fermion-2}) requires
\begin{align}
k\pm 2,\
k\pm 6,\
-k+3,\
-k-1,\
-k+7,\
-k-5
\not=0\ (\mbox{mod.}\ N).
\end{align}
Taking $k=0$ as an example, we have the requirement
\begin{align}
\pm 2,\
\pm 6,\
+3,\
-1,\
+7,\
-5
\not=0\ (\mbox{mod.}\ N).
\end{align}
This then leads to $N=4$ or $N\geq 8$.  Combining the two conditions, we need to take $N\geq 10$.  Thus, we have found another possible set of parameters: $(\ell,k,N)=(2,0,10)$.

In summary, to unify the three chiral generations of SM fermions without
the occurrence of exotic fermions, one should take 
$(N,\eta_{\Psi^1},\eta_{\Psi^2})=(9,\omega^{-1},\omega^{-4})$, 
$(10,\omega^{-2},1)$, $\dots$  In such cases, only the fermions in the
$({\bf 16,3})(1)$ representation of 
$SO(10)\times SU(3)\times U(1)$ have zero modes.

\section{6D and 4D gauge anomaly cancellation}
\label{Sec:Anomaly}

From the above discussions, it is clear that one choice of having
the three chiral generations of SM fermions is to take $(N,\eta_{\Psi^1},\eta_{\Psi^2})=(9,\omega^{-1},\omega^{-4})$.
In this case, we obtain the 4D left-handed fermions in $({\bf 16,3})(1)$ of
$SO(10)\times SU(3)\times U(1)$ 
from a single 6D fermion $\Psi_{\bf 128}^{\alpha=1}$ without having exotic
fermions.  In the following, we discuss 6D and 4D gauge anomaly cancellation in this scenario.

First, we consider the 6D $SO(16)$ gauge anomaly cancellation.  Our
model includes 6D Weyl fermions and thus can in principle induce 6D gauge anomalies.
By introducing one 6D positive Weyl fermion in {\bf 128} of $SO(16)$ and 
another 6D negative Weyl fermion in {\bf 128} of $SO(16)$, the fermions in the model are vectorlike in the 6D gauge theory.  Therefore, the 6D $SO(16)$ gauge anomaly is trivially canceled.

Next, we discuss how to cancel the 4D gauge anomalies.
As is well known, there are two types of 4D gauge anomalies: pure gauge
anomalies and mixed gauge anomalies.
Pure gauge anomalies are determined by the group structure in the case
of semi-simple Lie groups, and there are no 4D pure gauge anomalies
except for the $SU(n)$ and $U(1)$ groups
\cite{Banks1976,Okubo:1977sc,Patera:1981sc,Yamatsu:2015gut}.
In other words, in the present case the pure gauge quantum anomaly is
only in the $SU(3)$ and $U(1)$ symmetries, i.e., the
$SU(3)\mbox{-}SU(3)\mbox{-}SU(3)$ and 
$U(1)\mbox{-}U(1)\mbox{-}U(1)$ gauge anomalies.
There are only three mixed gauge anomalies, i.e., the
$SO(10)\mbox{-}SO(10)\mbox{-}U(1)$,
$SU(3)\mbox{-}SU(3)\mbox{-}U(1)$, and
$\mbox{grav}\mbox{-}\mbox{grav}\mbox{-}U(1)$ anomalies.
Furthermore, the pure gauge anomalies can be written in
terms of third-order Dynkin indices or anomaly coefficients
and the mixed gauge anomalies can be written in terms of second-order Dynkin
indices and $U(1)$ charges, as follows:
\begin{align}
&A_{\rm bulk}^{SU(3)\mbox{-}SU(3)\mbox{-}SU(3)}=
16\times A_{SU(3)}{(\bf 3)}=+16,
\nonumber\\
&A_{\rm bulk}^{U(1)\mbox{-}U(1)\mbox{-}U(1)}=
16\times 3\times (+1)^3=+48,
\nonumber\\
&A_{\rm bulk}^{SO(10)\mbox{-}SO(10)\mbox{-}U(1)}=
3\times T_{SO(10)}{(\bf 16)}\times (+1)=+6,
\nonumber\\
&A_{\rm bulk}^{SU(3)\mbox{-}SU(3)\mbox{-}U(1)}=
16\times T_{SU(3)}{(\bf 3)}\times (+1)=+8,
\nonumber\\
&A_{\rm bulk}^{\mbox{grav}\mbox{-}\mbox{grav}\mbox{-}U(1)}=
16\times 3\times T_{SO(4)}{(\bf 2,1)}\times (+1)=+24,
\label{Eq:SO16-gauge-anomaly-bulk}
\end{align}
where $A_{SU(3)}{(\bf 3)}=+1$, $T_{SO(10)}{(\bf 16)}=2$,
$T_{SU(3)}{(\bf 3)}=1/2$, and $T_{SO(4)}{({\bf 2,1})}=1/2$
\cite{Yamatsu:2015gut}.

We now examine what kind of 4D fields should be introduced to
cancel out the gauge anomaly generated by the 6D fields given in
Eq.~(\ref{Eq:SO16-gauge-anomaly-bulk}). 
It turns out to be impossible to uniquely determine the required set of 4D
fields solely by the requirement that the 4D gauge anomalies arising
from the 6D fields cancel out.  In the following, we provide an explicit example.
One way is to introduce 4D localized left-handed fermions in 
$({\bf 10,\overline{3}})(-1)$ and
$({\bf 1,3})(+1)$
of $SO(10)\times SU(3)\times U(1)$, denoted by
$\psi_{L({\bf 10,\overline{3}})(-1)}^{(\beta)}$ $(\beta=1,2)$ and
$\psi_{L({\bf 1,{3}})(+1)}^{(\gamma)}$ $(\gamma=1,2,3,4)$, respectively.
In the case, the $SO(10)\mbox{-}SO(10)\mbox{-}U(1)$ 
mixed gauge anomaly is canceled by 
$\psi_{L({\bf 10,\overline{3}})(-1)}^{(\beta)}$;
the remaining anomalies 
are canceled by the combination of $\psi_{L({\bf 10,\overline{3}})(-1)}^{(\beta)}$ and
$\psi_{L({\bf 1,{3}})(+1)}^{(\gamma)}$.
When we take into account the vacuum expectation values (VEVs) of
the $SU(3)$- and $U(1)$-breaking scalar fields,
these 4D fields obtain masses that do not appear at low energies
and will not be discussed further in this paper.

\section{Symmetry breaking via the VEVs of scalars}
\label{Sec:Symmetry-Breaking}

Next, we consider the scalar content to realize the SM gauge symmetry at 
low energies. The set of scalar fields, which properly break the
$SO(16)$ symmetry into $G_{\rm SM}$, is also not uniquely determined.
Here we provide a working example by introducing localized scalars 
$\phi_{{\bf (10,1)}(0)}$,
$\phi_{{\bf (45,1})(0)}$,
$\phi_{{\bf (16,1)}(0)}$,
$\phi_{{\bf (1,3)}(0)}$,
$\phi_{{\bf (1,6)}(+2)}$, and
$\phi_{{\bf (1,1)}(+1)}$
to realize $G_{\rm SM}$ at low energies.
In this scenario, the symmetry-breaking pattern is given as follows:
\begin{align}
\begin{array}{rll}
SO(16)\ \to&
SO(10)\times SU(3)\times U(1)
&\mbox{by the non-trivial space structure}\\
\to&
SO(10)
&\mbox{by the VEVs}\
\langle\phi_{{\bf (1,3)}(0)}\rangle,
\langle\phi_{{\bf (1,6)}(+2)}\rangle,
\langle\phi_{{\bf (1,1)}(+1)}\rangle
\not=0\\
\to&
G_{\rm SM}
&\mbox{by the VEVs}\
\langle\phi_{{\bf (45,1)}(0)}\rangle,
\langle\phi_{{\bf (16,1)}(0)}\rangle
\not=0\\
\to&
SU(3)_C\times U(1)_{\rm EM}
&\mbox{by the VEV}\
\langle\phi_{{\bf (10,1)}(0)}\rangle
\not=0,\\
\end{array}
\end{align}
where we assume that the Kaluza-Klein (KK) mass scales $1/R$ and the
non-vanishing VEVs are at the order of 
$M_{\rm GUT}\simeq O(10^{16}-10^{17})$\,GeV
except the VEV of the SM Higgs $\langle\phi_{{\bf (10,1)}(0)}\rangle$.
By assuming symmetry breaking with the scalar field VEVs given above, all
the introduced 4D fermions will attain large mass and will not appear at low
energies.  The $SO(10)$ GUT symmetry breaking to the SM gauge symmetry has been
thoroughly examined in, for example, 
Refs.~\cite{Pati:1974yy,Pati:1975ca,Mohapatra:1978fy,Slansky:1981yr,Yamatsu:2015gut}
and will not be discussed further here.

\section{Renormalization group equation for gauge coupling constant}
\label{Sec:RGE}

Here we analyze the renormalization group equation (RGE) evolution of the $SO(16)$ gauge coupling
constant at energy scales above the KK scale by using a
KK-mode expanded 4D effective theory as discussed in 
Ref.~\cite{Yamatsu:2015rge}.
Since the 4D fermions and scalar bosons do not have KK modes, they do not
change the qualitative properties of the RGE evolution above the KK
scale. In the following discussions, we will focus on the RGE evolution
affected by the KK particles from the 6D fields.

We first summarize some basic results of the RGEs for 4D gauge coupling
constants. The RGE for the $SO(16)$ gauge coupling constant $g_{SO(16)}$
is given in, e.g., 
Refs.~\cite{Slansky:1981yr,Yamatsu:2015gut,Yamatsu:2015rge}.
The RGE for the $SO(16)$ fine structure constant 
$\alpha_{SO(16)}(\mu):= g_{SO(16)}^2(\mu)/4\pi$
at energy scale $\mu$ is given by
\begin{align}
\frac{d}{d\mbox{log}(\mu)}(\alpha_{SO(16)}(\mu))^{-1}=
-\frac{b_{SO(16)}(\mu)}{2\pi}.
\label{Eq:RGE-gauge-coupling-4D-alpha}
\end{align}
where $b_{SO(16)}(\mu)$ is the $\beta$-function coefficient of $SO(16)$
at the energy scale $\mu$.  More explicitly,
\begin{align}
b_{SO(16)}(\mu)=
-\frac{11}{3}\sum_{\rm Vector}T_{SO(16)}(R_V)
+\frac{2}{3}\sum_{\rm Weyl}T_{SO(16)}(R_F)
+\frac{1}{6}\sum_{\rm Real}T_{SO(16)}(R_S).
\label{Eq:beta-function-coeff-general}
\end{align}
where Vector, Weyl, and Real under the summation symbol stand respectively for real vector, Weyl fermion, and real scalar fields in terms of 4D theories.
The summation is done for particles with mass smaller than the specified energy
scale $\mu$.
The vector bosons are the gauge bosons and thus belong to the
adjoint representation {\bf 120} of $SO(16)$ with $T(R_V={\bf 120})=C_2(SO(16))=14$, where
$C_2(SO(16))$ is the quadratic Casimir invariant of the adjoint
representation of $SO(16)$, and  
$T(R_X)$ is the Dynkin index of the irreducible representation $R_X$ of
$SO(16)$.  Finally, $T({\bf 128})=16$
\cite{Yamatsu:2015gut}.

Let us now consider the RGE for the 4D gauge coupling constant in the 6D gauge
theories given in Eq.~(\ref{Eq:RGE-gauge-coupling-4D-alpha})
by using the $\beta$ function coefficient given in
Eq.~(\ref{Eq:beta-function-coeff-general}), which depends on its
field content at the energy scale $\mu$.
For the $\beta$ function coefficient, we take into account the contribution from not only the zero modes but also the KK modes with masses less than the renormalization scale $\mu$.  Once specifying the mass spectrum of the model, we can calculate the RGE for the gauge coupling constant at one-loop level.
In general, it is difficult to write down the exact mass spectrum because it
depends on orbifold boundary conditions as well as the parameters of 6D bulk and 4D localized terms.  At the leading-order approximation, the masses of zero modes and 4D fields are much less than $\mu$ and those of the $k$-th KK modes can be set as $m=km_{KK}$.
By using this mass spectrum, the RGE of the gauge coupling constant
can be divided into two regimes: 
\begin{align}
\frac{d}{d\mbox{log}(\mu)}
\left(\alpha_{SO(16)}(\mu)\right)^{-1}\simeq
\left\{
\begin{array}{ll}
-\frac{1}{2\pi}b_{SO(16)}^{0}&\ \ \mbox{for}\ \ \mu<m_{KK}\\
-\frac{1}{2\pi}
\left(b_{SO(16)}^0+k\Delta b_{SO(16)}^{KK}\right)&\ \ \mbox{for}\ \
km_{KK}\leq\mu<\left(k+1\right)m_{KK}\\
\end{array}
\right.,
\label{Eq:RGE-4D-gauge-coupling-in-6D}
\end{align}
where $b_{SO(16)}^0$ is the $\beta$-function coefficient derived from
the 6D zero modes and 4D fields and $\Delta b_{SO(16)}^{KK}$ is the additional $\beta$-function coefficient induced by the KK modes of all 6D fields, both of which can be 
calculated by using Eq.~(\ref{Eq:beta-function-coeff-general}).
More explicitly, the latter is given by
\begin{align}
\Delta b_{SO(16)}^{KK}&=
-\frac{10}{3}C_2(SO(16))+\frac{4}{3}\sum_{\rm 6D\ Weyl}T_{SO(16)}(R_F)
\label{Eq:beta-function-coeff-KK-modes}
\end{align}
because each 6D bulk gauge field is decomposed into one 4D gauge boson and
two scalars and each 6D bulk Weyl fermion is decomposed into a 4D
Dirac fermion.
Our model includes one 6D $SO(16)$ gauge boson 
in the adjoint representation ${\bf 120}$ of $SO(16)$
and two 6D Weyl fermions in the spinor representation ${\bf 128}$ of $SO(16)$.
Therefore, Eq.~(\ref{Eq:beta-function-coeff-KK-modes}) gives 
\begin{align}
\Delta b_{SO(16)}^{KK}&=-\frac{10}{3}\times 14
+\frac{4}{3}\times 2 \times 16
=-4<0.
\end{align}
Therefore, the $SO(16)$ gauge coupling constant is found to be asymptotically free.

Finally, we comment on the RGE evolutions of the gauge coupling constants at
energy scales below the unification scale.  At energy scales below the KK
scale, the zero modes of the 6D fields and the 4D fields contribute to the
RGE evolutions of the gauge coupling constants.  As is well known, the
contribution of the SM particles alone does not unify the three SM gauge
coupling constants at a single high energy scale.
Therefore, non-SM particles of a specific mass spectrum or
symmetry breaking at scales between the electroweak and unification
scales are required to satisfactorily unify the gauge coupling
constants.  Related discussions can be found in various models 
\cite{Amaldi:1991zx,Deshpande:1992au,Deshpande:1992em,Chkareuli:1994ng,Babu:2015bna,Ferrari:2018rey,Abe:2021byq,Chiang:2023omu}.

\section{Summary and discussions}
\label{Sec:Summary}

In this paper, we have proposed a unified model for the SM gauge symmetry
and family symmetry based on the $SO(16)$ gauge symmetry on the 6D
spacetime $M^{4}\times D^2/\mathbb{Z}_N$ $(N\geq 9)$.
In this model, the three chiral generations of quarks and leptons in the SM can be
unified into one 6D Weyl fermion in the spinor representation of
$SO(16)$ when considering the symmetry-breaking pattern  
$SO(16)\to SO(10)\times SU(3)\times U(1)$.
The 6D $SO(16)$ gauge anomaly is canceled due to the vectorlike
nature of the Weyl fermion in the model, and the 4D gauge anomalies are canceled by introducing suitable 4D localized fermions at the fixed point. 
The gauge coupling constant of $SO(16)$ is found to be asymptotically free above the KK scale.

We note in passing that an $SO(18)$ model can unify the 6D Weyl fermions
into the spinor representation {\bf 256} with three chiral
generations of the SM fermions without 6D $SO(18)$ gauge anomaly.
Using the same arguments for $SO(16)$ given in Sec.~\ref{Sec:RGE}, the
equation corresponding to Eq.~(\ref{Eq:beta-function-coeff-KK-modes}) in such a model is
given by 
\begin{align}
\Delta b_{SO(18)}^{KK}&=
-\frac{10}{3}C_2(SO(18))+\frac{4}{3}\sum_{\rm 6D\ Weyl}T_{SO(18)}(R_F)
=-\frac{10}{3}\times 16+\frac{4}{3}\times 2 \times 32=+32>0,
\end{align}
where $C_2(SO(18))=16$ and $T_{SO(18)}({\bf 256})=32$ 
\cite{Yamatsu:2015gut}.  Hence, the gauge coupling constant in the $SO(18)$ model is not
asymptotically free.  Basically, the cause of this difference arises from the fact that the
dimensions of the adjoint and spinor representations increase with increasing $N$ in $SO(N)$.

It is worth mentioning that for gauge-Higgs unification (GHU) models, which unify
gauge bosons and Higgs bosons
\cite{Hosotani:1983xw,Hosotani:1988bm,Davies:1987ei,Davies:1988wt,Hatanaka:1998yp},
one may use the $SO(N)$ gauge symmetry as in this work to construct a model that embeds
$SU(3)_C\times SO(5)_W\times U(1)$ electroweak GHU models
\cite{Agashe:2004rs,Hosotani:2008tx,Funatsu:2019xwr}
or $SO(11)$ gauge-Higgs GUT models
\cite{Hosotani:2015hoa,Furui:2016owe,Hosotani:2017ghg,Hosotani:2017edv,Englert:2019xhz,Englert:2020eep}.
For phenomenological aspects of the models, see Refs.~\cite{Funatsu:2013ni,Funatsu:2019fry,Funatsu:2020znj,Funatsu:2020haj,Funatsu:2021gnh,Funatsu:2021yrh,Funatsu:2022spb,Funatsu:2023jng,Hosotani:2023yhs,Yamatsu:2023bde,Hosotani:2023poh,Irles:2024ipg}.

Finally, we make a brief comment on the masses of quarks and leptons.
The analysis of mass matrices in models with $SU(3)$ family symmetry has already
been done in
Refs.~\cite{Koide:1987vp,Koide:1995xk,Koide:2008ey,Koide:2013eca,Koide:2014doa,Koide:2019bgk,King:2001uz,Ross:2002fb,King:2003rf,Appelquist:2006ag,Appelquist:2006qq,Yang:2011qj,Wu:2012ria}.
Though phenomenologically important, a detailed study of masses and mixing in this model, however, is outside the scope of this paper and will be discussed in future work.

\section*{Acknowledgments}

This work was supported in part 
by the Japan Society for the Promotion of Science (JSPS) KAKENHI Grant
No.~JP21H05182 (N.Y.),
by the National Science and Technology Council of Taiwan under Grant
Nos.~NSTC-111-2112-M-002-018-MY3 (C.W.C.),
and
by the National Key Research and Development Program of China Grant
No. 2020YFC2201504, by the Project No. 12275333 supported by the
National Natural Science Foundation of China, by the Scientific
Instrument Developing Project of the Chinese Academy of Sciences, Grant
No. YJKYYQ20190049, and by the International Partnership Program of
Chinese Academy of Sciences for Grand Challenges, Grant
No. 112311KYSB20210012 (T.L.).



\bibliographystyle{utphys} 
\bibliography{../../arxiv/reference}

\providecommand{\href}[2]{#2}\begingroup\raggedright\begin{thebibliography}{10}

\bibitem{Wilczek:1978xi}
F.~Wilczek and A.~Zee, ``{Horizontal Interaction and Weak Mixing Angles},''
\href{http://dx.doi.org/10.1103/PhysRevLett.42.421}{{ Phys. Rev. Lett.}
  {\bfseries 42} (1979) 421}.

\bibitem{Froggatt:1978nt}
C.~D. Froggatt and H.~B. Nielsen, ``{Hierarchy of Quark Masses, Cabibbo Angles
  and CP Violation},''
\href{http://dx.doi.org/10.1016/0550-3213(79)90316-X}{{ Nucl. Phys.} {\bfseries
  B147} (1979) 277}.

\bibitem{Yanagida:1979as}
T.~Yanagida, ``{Horizontal Gauge Symmetry and Masses of Neutrinos},''. In
  Proceedings of the Workshop on the Baryon Number of the Universe and Unified
  Theories, Tsukuba, Japan, p95 (1979).

\bibitem{Maehara:1979kf}
T.~Maehara and T.~Yanagida, ``{Gauge Symmetry of Horizontal Flavor},''
\href{http://dx.doi.org/10.1143/PTP.61.1434}{{ Prog. Theor. Phys.} {\bfseries
  61} (1979) 1434}.

\bibitem{Inoue:1994qz}
K.~Inoue, ``{Generations of Quarks and Leptons from Noncompact Horizontal
  Symmetry},'' \href{http://dx.doi.org/10.1143/PTP.93.403}{{ Prog. Theor.
  Phys.} {\bfseries 93} (1995) 403--416},
\href{http://arxiv.org/abs/hep-ph/9410220}{{\ttfamily arXiv:hep-ph/9410220}}.

\bibitem{King:2001uz}
S.~F. King and G.~G. Ross, ``{Fermion Masses and Mixing Angles from $SU(3)$
  Family Symmetry},'' \href{http://dx.doi.org/10.1016/S0370-2693(01)01139-X}{{
  Phys. Lett.} {\bfseries B520} (2001) 243--253},
\href{http://arxiv.org/abs/hep-ph/0108112}{{\ttfamily arXiv:hep-ph/0108112}}.

\bibitem{Maekawa2004}
N.~Maekawa and T.~Yamashita, ``{Horizontal Symmetry in Higgs Sector of GUT with
  $U(1)_A$ Symmetry},''
  \href{http://dx.doi.org/10.1088/1126-6708/2004/07/009}{{ JHEP} {\bfseries 07}
  (2004) 009},
\href{http://arxiv.org/abs/hep-ph/0404020}{{\ttfamily arXiv:hep-ph/0404020}}.

\bibitem{Yamatsu:2007}
K.~Inoue and N.~Yamatsu, ``{Charged Lepton and Down-Type Quark Masses in
  $SU(1,1)$ Model and the Structure of Higgs Sector},''
  \href{http://dx.doi.org/10.1143/PTP.119.775}{{ Prog. Theor. Phys.} {\bfseries
  119} (2008) 775--796},
\href{http://arxiv.org/abs/0712.2938}{{\ttfamily arXiv:0712.2938 [hep-ph]}}.

\bibitem{Yamatsu:2012}
N.~Yamatsu, ``{New Mixing Structures of Chiral Generations in a Model with
  Noncompact Horizontal Symmetry},''
  \href{http://dx.doi.org/10.1093/ptep/pts079}{{ Prog. Theor. Exp. Phys.}
  {\bfseries 2013} (2013) 023B03},
\href{http://arxiv.org/abs/1209.6318}{{\ttfamily arXiv:1209.6318 [hep-ph]}}.

\bibitem{Yoshioka:1999ds}
K.~Yoshioka, ``{On Fermion Mass Hierarchy with Extra Dimensions},''
  \href{http://dx.doi.org/10.1142/S0217732300000062,
  10.1016/S0217-7323(00)00006-2}{{ Mod. Phys. Lett.} {\bfseries A15} (2000)
  29--40},
\href{http://arxiv.org/abs/hep-ph/9904433}{{\ttfamily arXiv:hep-ph/9904433
  [hep-ph]}}.

\bibitem{Fujimoto:2012wv}
Y.~Fujimoto, T.~Nagasawa, K.~Nishiwaki, and M.~Sakamoto, ``{Quark Mass
  Hierarchy and Mixing via Geometry of Extra Dimension with Point
  Interactions},'' \href{http://dx.doi.org/10.1093/ptep/pts097}{{ PTEP}
  {\bfseries 2013} (2013) 023B07},
\href{http://arxiv.org/abs/1209.5150}{{\ttfamily arXiv:1209.5150 [hep-ph]}}.

\bibitem{Georgi:1974sy}
H.~Georgi and S.~L. Glashow, ``{Unity of All Elementary Particle Forces},''
\href{http://dx.doi.org/10.1103/PhysRevLett.32.438}{{ Phys. Rev. Lett.}
  {\bfseries 32} (1974) 438--441}.

\bibitem{Inoue:1977qd}
K.~Inoue, A.~Kakuto, and Y.~Nakano, ``{Unification of the Lepton-Quark World by
  the Gauge Group SU(6)},''
\href{http://dx.doi.org/10.1143/PTP.58.630}{{ Prog.Theor.Phys.} {\bfseries 58}
  (1977) 630}.

\bibitem{Fritzsch:1974nn}
H.~Fritzsch and P.~Minkowski, ``{Unified Interactions of Leptons and
  Hadrons},''
\href{http://dx.doi.org/10.1016/0003-4916(75)90211-0}{{ Ann. Phys.} {\bfseries
  93} (1975) 193--266}.

\bibitem{Ida:1980ea}
M.~Ida, Y.~Kayama, and T.~Kitazoe, ``{Inclusion of Generations in SO(14)},''
\href{http://dx.doi.org/10.1143/PTP.64.1745}{{ Prog. Theor. Phys.} {\bfseries
  64} (1980) 1745}.

\bibitem{Fujimoto:1981bv}
Y.~Fujimoto, ``{SO(18) Unification},''
\href{http://dx.doi.org/10.1103/PhysRevD.26.3183}{{ Phys. Rev.} {\bfseries D26}
  (1982) 3183}.

\bibitem{Gursey:1975ki}
F.~Gursey, P.~Ramond, and P.~Sikivie, ``{A Universal Gauge Theory Model Based
  on $E_6$},''
\href{http://dx.doi.org/10.1016/0370-2693(76)90417-2}{{ Phys. Lett.} {\bfseries
  B60} (1976) 177}.

\bibitem{Kawamura:1999nj}
Y.~Kawamura, ``{Gauge Symmetry Breaking from Extra Space $S^1/Z_2$},''
  \href{http://dx.doi.org/10.1143/PTP.103.613}{{ Prog. Theor. Phys.} {\bfseries
  103} (2000) 613--619},
\href{http://arxiv.org/abs/hep-ph/9902423}{{\ttfamily arXiv:hep-ph/9902423
  [hep-ph]}}.

\bibitem{Kawamura:2000ir}
Y.~Kawamura, ``{Split Multiplets, Coupling Unification and Extra Dimension},''
  \href{http://dx.doi.org/10.1143/PTP.105.691}{{ Prog. Theor. Phys.} {\bfseries
  105} (2001) 691--696},
\href{http://arxiv.org/abs/hep-ph/0012352}{{\ttfamily arXiv:hep-ph/0012352}}.

\bibitem{Kawamura:2000ev}
Y.~Kawamura, ``{Triplet-Doublet Splitting, Proton Stability and Extra
  Dimension},'' \href{http://dx.doi.org/10.1143/PTP.105.999}{{ Prog. Theor.
  Phys.} {\bfseries 105} (2001) 999--1006},
\href{http://arxiv.org/abs/hep-ph/0012125}{{\ttfamily arXiv:hep-ph/0012125}}.

\bibitem{Hall:2001pg}
L.~J. Hall and Y.~Nomura, ``{Gauge Unification in Higher Dimensions},''
  \href{http://dx.doi.org/10.1103/PhysRevD.64.055003}{{ Phys.Rev.} {\bfseries
  D64} (2001) 055003},
\href{http://arxiv.org/abs/hep-ph/0103125}{{\ttfamily arXiv:hep-ph/0103125
  [hep-ph]}}.

\bibitem{Burdman:2002se}
G.~Burdman and Y.~Nomura, ``{Unification of Higgs and Gauge Fields in
  Five-Dimensions},'' \href{http://dx.doi.org/10.1016/S0550-3213(03)00088-9}{{
  Nucl. Phys.} {\bfseries B656} (2003) 3--22},
\href{http://arxiv.org/abs/hep-ph/0210257}{{\ttfamily arXiv:hep-ph/0210257
  [hep-ph]}}.

\bibitem{Lim:2007jv}
C.~S. Lim and N.~Maru, ``{Towards a Realistic Grand Gauge-Higgs Unification},''
  \href{http://dx.doi.org/10.1016/j.physletb.2007.07.053}{{ Phys.Lett.}
  {\bfseries B653} (2007) 320--324},
\href{http://arxiv.org/abs/0706.1397}{{\ttfamily arXiv:0706.1397 [hep-ph]}}.

\bibitem{Kim:2002im}
H.~D. Kim and S.~Raby, ``{Unification in 5-D SO(10)},''
  \href{http://dx.doi.org/10.1088/1126-6708/2003/01/056}{{ JHEP} {\bfseries 01}
  (2003) 056},
\href{http://arxiv.org/abs/hep-ph/0212348}{{\ttfamily arXiv:hep-ph/0212348
  [hep-ph]}}.

\bibitem{Fukuyama:2008pw}
T.~Fukuyama and N.~Okada, ``{A Simple SO(10) GUT in Five Dimensions},''
  \href{http://dx.doi.org/10.1103/PhysRevD.78.015005}{{ Phys. Rev.} {\bfseries
  D78} (2008) 015005},
\href{http://arxiv.org/abs/0803.1758}{{\ttfamily arXiv:0803.1758 [hep-ph]}}.

\bibitem{Hosotani:2015hoa}
Y.~Hosotani and N.~Yamatsu, ``{Gauge-Higgs Grand Unification},''
  \href{http://dx.doi.org/10.1093/ptep/ptv153}{{ Prog. Theor. Exp. Phys.}
  {\bfseries 2015} (2015) 111B01},
\href{http://arxiv.org/abs/1504.03817}{{\ttfamily arXiv:1504.03817 [hep-ph]}}.

\bibitem{Ramond:1979py}
P.~Ramond, ``{The Family Group in Grand Unified Theories},'' in {
  {International Symposium on Fundamentals of Quantum Theory and Quantum Field
  Theory}}, pp.~265--280.
\newblock 1979.
\newblock
\href{http://arxiv.org/abs/hep-ph/9809459}{{\ttfamily arXiv:hep-ph/9809459
  [hep-ph]}}.
\newblock

\bibitem{Arkani-Hamed:2001vvu}
N.~Arkani-Hamed, T.~Gregoire, and J.~G. Wacker, ``{Higher Dimensional
  Supersymmetry in 4-D Superspace},''
  \href{http://dx.doi.org/10.1088/1126-6708/2002/03/055}{{ JHEP} {\bfseries 03}
  (2002) 055}, \href{http://arxiv.org/abs/hep-th/0101233}{{\ttfamily
  arXiv:hep-th/0101233}}.

\bibitem{Kawamura:2007cm}
Y.~Kawamura, T.~Kinami, and K.-y. Oda, ``{Orbifold Family Unification},''
  \href{http://dx.doi.org/10.1103/PhysRevD.76.035001}{{ Phys. Rev.} {\bfseries
  D76} (2007) 035001},
\href{http://arxiv.org/abs/hep-ph/0703195}{{\ttfamily arXiv:hep-ph/0703195}}.

\bibitem{Kawamura:2009gr}
Y.~Kawamura and T.~Miura, ``{Orbifold Family Unification in $SO(2N)$ Gauge
  Theory},'' \href{http://dx.doi.org/10.1103/PhysRevD.81.075011}{{ Phys. Rev.}
  {\bfseries D81} (2010) 075011},
\href{http://arxiv.org/abs/0912.0776}{{\ttfamily arXiv:0912.0776 [hep-ph]}}.

\bibitem{Goto:2013jma}
Y.~Goto, Y.~Kawamura, and T.~Miura, ``{Orbifold Family Unification on Six
  Dimensions},'' \href{http://dx.doi.org/10.1103/PhysRevD.88.055016}{{
  Phys.Rev.} {\bfseries D88} no.~5, (2013) 055016},
\href{http://arxiv.org/abs/1307.2631}{{\ttfamily arXiv:1307.2631}}.

\bibitem{Fonseca:2015aoa}
R.~M. Fonseca, ``{On the Chirality of the SM and the Fermion Content of
  GUTs},'' \href{http://dx.doi.org/10.1016/j.nuclphysb.2015.06.012}{{ Nucl.
  Phys.} {\bfseries B897} (2015) 757--780},
\href{http://arxiv.org/abs/1504.03695}{{\ttfamily arXiv:1504.03695 [hep-ph]}}.

\bibitem{Albright:2016lpi}
C.~H. Albright, R.~P. Feger, and T.~W. Kephart, ``{Unification of Gauge,
  Family, and Flavor Symmetries Illustrated in Gauged $SU(12)$ Models},''
  \href{http://dx.doi.org/10.1103/PhysRevD.93.075032}{{ Phys. Rev.} {\bfseries
  D93} no.~7, (2016) 075032},
\href{http://arxiv.org/abs/1601.07523}{{\ttfamily arXiv:1601.07523 [hep-ph]}}.

\bibitem{Goto:2017zsx}
Y.~Goto and Y.~Kawamura, ``{Orbifold Family Unification Using Vectorlike
  Representation on Six Dimensions},''
\href{http://arxiv.org/abs/1712.06444}{{\ttfamily arXiv:1712.06444 [hep-ph]}}.

\bibitem{Reig:2017nrz}
M.~Reig, J.~W.~F. Valle, C.~A. Vaquera-Araujo, and F.~Wilczek, ``{A Model of
  Comprehensive Unification},''
  \href{http://dx.doi.org/10.1016/j.physletb.2017.10.038}{{ Phys. Lett.}
  {\bfseries B774} (2017) 667--670},
\href{http://arxiv.org/abs/1706.03116}{{\ttfamily arXiv:1706.03116 [hep-ph]}}.

\bibitem{Yamatsu:2018fsg}
N.~Yamatsu, ``{Family Unification in Special Grand Unification},''
  \href{http://dx.doi.org/10.1093/ptep/pty100}{{ Prog. Theor. Exp. Phys.}
  {\bfseries 2018} no.~9, (2018) 091B01},
\href{http://arxiv.org/abs/1807.10855}{{\ttfamily arXiv:1807.10855 [hep-ph]}}.

\bibitem{Ekstedt:2020gaj}
A.~Ekstedt, R.~M. Fonseca, and M.~Malinsk\'y, ``{Flavorgenesis in an SU(19)
  model},'' \href{http://dx.doi.org/10.1016/j.physletb.2021.136212}{{ Phys.
  Lett. B} {\bfseries 816} (2021) 136212},
  \href{http://arxiv.org/abs/2009.03909}{{\ttfamily arXiv:2009.03909
  [hep-ph]}}.

\bibitem{Maru:2024ghd}
N.~Maru and R.~Nago, ``{Attempt at Constructing a Model of Grand Gauge-Higgs
  Unification with Family Unification},''
  \href{http://dx.doi.org/10.1103/PhysRevD.109.115005}{{ Phys. Rev. D}
  {\bfseries 109} no.~11, (2024) 115005},
  \href{http://arxiv.org/abs/2403.02731}{{\ttfamily arXiv:2403.02731
  [hep-ph]}}.

\bibitem{Yamatsu:2015gut}
N.~Yamatsu, ``{Finite-Dimensional Lie Algebras and Their Representations for
  Unified Model Building},''
\href{http://arxiv.org/abs/1511.08771}{{\ttfamily arXiv:1511.08771 [hep-ph]}}.

\bibitem{Yamatsu:2017sgu}
N.~Yamatsu, ``{Special Grand Unification},''
  \href{http://dx.doi.org/10.1093/ptep/ptx088}{{ Prog. Theor. Exp. Phys.}
  {\bfseries 2017} no.~6, (2017) 061B01},
\href{http://arxiv.org/abs/1704.08827}{{\ttfamily arXiv:1704.08827 [hep-ph]}}.

\bibitem{Yamatsu:2017ssg}
N.~Yamatsu, ``{String-Inspired Special Grand Unification},''
  \href{http://dx.doi.org/10.1093/ptep/ptx135}{{ Prog. Theor. Exp. Phys.}
  {\bfseries 2017} no.~10, (2017) 101B01},
\href{http://arxiv.org/abs/1708.02078}{{\ttfamily arXiv:1708.02078 [hep-ph]}}.

\bibitem{Li:2001dt}
T.-j. Li, ``{Discrete Symmetry and GUT Breaking},''
  \href{http://dx.doi.org/10.1007/s10052-002-0968-0}{{ Eur. Phys. J. C}
  {\bfseries 24} (2002) 595--612},
  \href{http://arxiv.org/abs/hep-th/0110065}{{\ttfamily arXiv:hep-th/0110065}}.

\bibitem{Huang:2004ui}
C.-S. Huang, J.~Jiang, and T.~Li, ``{Supersymmetric SO(10) Models Inspired by
  Deconstruction},'' \href{http://dx.doi.org/10.1016/j.nuclphysb.2004.09.023}{{
  Nucl. Phys. B} {\bfseries 702} (2004) 109--130},
  \href{http://arxiv.org/abs/hep-ph/0407180}{{\ttfamily arXiv:hep-ph/0407180}}.

\bibitem{Banks1976}
J.~Banks and H.~Georgi, ``{Comment on Gauge Theories Without Anomalies},''
\href{http://dx.doi.org/10.1103/PhysRevD.14.1159}{{ Phys. Rev.} {\bfseries D14}
  (1976) 1159--1160}.

\bibitem{Okubo:1977sc}
S.~Okubo, ``{Gauge Groups Without Triangular Anomaly},''
\href{http://dx.doi.org/10.1103/PhysRevD.16.3528}{{ Phys.Rev.} {\bfseries D16}
  (1977) 3528}.

\bibitem{Patera:1981sc}
J.~Patera and R.~T. Sharp, ``{On the Triangle Anomaly Number of $SU(n)$
  Representaions},''
\href{http://dx.doi.org/10.1063/1.524815}{{ J. Math. Phys.} {\bfseries 22}
  (1981) 2352}.

\bibitem{Pati:1974yy}
J.~C. Pati and A.~Salam, ``{Lepton Number as the Fourth Color},''
\href{http://dx.doi.org/10.1103/PhysRevD.10.275}{{ Phys. Rev.} {\bfseries D10}
  (1974) 275--289}.

\bibitem{Pati:1975ca}
J.~Pati, A.~Salam, and J.~Strathdee, ``{On Fermion number and its
  conservation},'' \href{http://dx.doi.org/10.1007/BF02849600}{{ Nuovo Cim. A}
  {\bfseries 26} (1975) 72--83}.

\bibitem{Mohapatra:1978fy}
R.~N. Mohapatra and G.~Senjanovic, ``{Natural Suppression of Strong P and T
  Noninvariance},''
\href{http://dx.doi.org/10.1016/0370-2693(78)90243-5}{{ Phys. Lett.} {\bfseries
  B79} (1978) 283--286}.

\bibitem{Slansky:1981yr}
R.~Slansky, ``{Group Theory for Unified Model Building},''
\href{http://dx.doi.org/10.1016/0370-1573(81)90092-2}{{ Phys. Rept.} {\bfseries
  79} (1981) 1--128}.

\bibitem{Yamatsu:2015rge}
N.~Yamatsu, ``{Gauge Coupling Unification in Gauge-Higgs Grand Unification},''
  \href{http://dx.doi.org/10.1093/ptep/ptw023}{{ Prog. Theor. Exp. Phys.}
  {\bfseries 2016} (2016) 043B02},
\href{http://arxiv.org/abs/1512.05559}{{\ttfamily arXiv:1512.05559 [hep-ph]}}.

\bibitem{Amaldi:1991zx}
U.~Amaldi, W.~de~Boer, P.~H. Frampton, H.~Furstenau, and J.~T. Liu,
  ``{Consistency Checks of Grand Unified Theories},''
  \href{http://dx.doi.org/10.1016/0370-2693(92)91158-6}{{ Phys. Lett. B}
  {\bfseries 281} (1992) 374--382}.

\bibitem{Deshpande:1992au}
N.~Deshpande, E.~Keith, and P.~B. Pal, ``{Implications of LEP Results for
  SO(10) Grand Unification},''
\href{http://dx.doi.org/10.1103/PhysRevD.46.2261}{{ Phys.Rev.} {\bfseries D46}
  (1992) 2261--2264}.

\bibitem{Deshpande:1992em}
N.~Deshpande, E.~Keith, and P.~B. Pal, ``{Implications of LEP Results for
  SO(10) Grand Unification with Two Intermediate Stages},''
  \href{http://dx.doi.org/10.1103/PhysRevD.47.2892}{{ Phys.Rev.} {\bfseries
  D47} (1993) 2892--2896},
\href{http://arxiv.org/abs/hep-ph/9211232}{{\ttfamily arXiv:hep-ph/9211232
  [hep-ph]}}.

\bibitem{Chkareuli:1994ng}
J.~L. Chkareuli, I.~G. Gogoladze, and A.~B. Kobakhidze, ``{Natural Non-SUSY
  SU(N) GUTs},'' \href{http://dx.doi.org/10.1016/0370-2693(94)91298-X}{{ Phys.
  Lett. B} {\bfseries 340} (1994) 63--66}.

\bibitem{Babu:2015bna}
K.~S. Babu and S.~Khan, ``{Minimal Nonsupersymmetric $SO(10)$ Model: Gauge
  Coupling Unification, Proton Decay, and Fermion Masses},''
  \href{http://dx.doi.org/10.1103/PhysRevD.92.075018}{{ Phys. Rev. D}
  {\bfseries 92} no.~7, (2015) 075018},
  \href{http://arxiv.org/abs/1507.06712}{{\ttfamily arXiv:1507.06712
  [hep-ph]}}.

\bibitem{Ferrari:2018rey}
S.~Ferrari, T.~Hambye, J.~Heeck, and M.~H. Tytgat, ``{$SO(10)$ Paths to Dark
  Matter},'' \href{http://dx.doi.org/10.1103/PhysRevD.99.055032}{{ Phys. Rev.
  D} {\bfseries 99} no.~5, (2019) 055032},
  \href{http://arxiv.org/abs/1811.07910}{{\ttfamily arXiv:1811.07910
  [hep-ph]}}.

\bibitem{Abe:2021byq}
Y.~Abe, T.~Toma, K.~Tsumura, and N.~Yamatsu, ``{Pseudo-Nambu-Goldstone Dark
  Matter Model Inspired by Grand Unification},''
  \href{http://dx.doi.org/10.1103/PhysRevD.104.035011}{{ Phys. Rev. D}
  {\bfseries 104} (2021) 035011},
  \href{http://arxiv.org/abs/2104.13523}{{\ttfamily arXiv:2104.13523
  [hep-ph]}}.

\bibitem{Chiang:2023omu}
C.-W. Chiang, K.~Tsumura, Y.~Uchida, and N.~Yamatsu, ``{Pseudo-Nambu-Goldstone
  Dark Matter in SU(7) Grand Unification},''
  \href{http://dx.doi.org/10.1103/PhysRevD.109.055040}{{ Phys. Rev. D}
  {\bfseries 109} no.~5, (2024) 055040},
  \href{http://arxiv.org/abs/2311.13753}{{\ttfamily arXiv:2311.13753
  [hep-ph]}}.

\bibitem{Hosotani:1983xw}
Y.~Hosotani, ``{Dynamical Mass Generation by Compact Extra Dimensions},''
\href{http://dx.doi.org/10.1016/0370-2693(83)90170-3}{{ Phys.Lett.} {\bfseries
  B126} (1983) 309}.

\bibitem{Hosotani:1988bm}
Y.~Hosotani, ``{Dynamics of Nonintegrable Phases and Gauge Symmetry
  Breaking},''
\href{http://dx.doi.org/10.1016/0003-4916(89)90015-8}{{ Annals Phys.}
  {\bfseries 190} (1989) 233}.

\bibitem{Davies:1987ei}
A.~T. Davies and A.~McLachlan, ``{Gauge Group Breaking By Wilson Loops},''
\href{http://dx.doi.org/10.1016/0370-2693(88)90776-9}{{ Phys. Lett.} {\bfseries
  B200} (1988) 305}.

\bibitem{Davies:1988wt}
A.~T. Davies and A.~McLachlan, ``{Congruency Class Effects in the Hosotani
  Model},''
\href{http://dx.doi.org/10.1016/0550-3213(89)90569-5}{{ Nucl. Phys.} {\bfseries
  B317} (1989) 237}.

\bibitem{Hatanaka:1998yp}
H.~Hatanaka, T.~Inami, and C.~S. Lim, ``{The Gauge Hierarchy Problem and Higher
  Dimensional Gauge Theories},''
  \href{http://dx.doi.org/10.1142/S021773239800276X}{{ Mod. Phys. Lett.}
  {\bfseries A13} (1998) 2601--2612},
\href{http://arxiv.org/abs/hep-th/9805067}{{\ttfamily arXiv:hep-th/9805067}}.

\bibitem{Agashe:2004rs}
K.~Agashe, R.~Contino, and A.~Pomarol, ``{The Minimal Composite Higgs Model},''
  \href{http://dx.doi.org/10.1016/j.nuclphysb.2005.04.035}{{ Nucl. Phys.}
  {\bfseries B719} (2005) 165--187},
\href{http://arxiv.org/abs/hep-ph/0412089}{{\ttfamily arXiv:hep-ph/0412089
  [hep-ph]}}.

\bibitem{Hosotani:2008tx}
Y.~Hosotani, K.~Oda, T.~Ohnuma, and Y.~Sakamura, ``{Dynamical Electroweak
  Symmetry Breaking in $SO(5) \times U(1)$ Gauge-Higgs Unification with Top and
  Bottom Quarks},'' \href{http://dx.doi.org/10.1103/PhysRevD.78.096002,
  10.1103/PhysRevD.79.079902}{{ Phys.Rev.} {\bfseries D78} (2008) 096002},
\href{http://arxiv.org/abs/0806.0480}{{\ttfamily arXiv:0806.0480 [hep-ph]}}.

\bibitem{Funatsu:2019xwr}
S.~Funatsu, H.~Hatanaka, Y.~Hosotani, Y.~Orikasa, and N.~Yamatsu, ``{GUT
  Inspired $SO(5) \times U(1) \times SU(3)$ Gauge-Higgs Unification},''
  \href{http://dx.doi.org/10.1103/PhysRevD.99.095010}{{ Phys. Rev. D}
  {\bfseries 99} (2019) 095010},
\href{http://arxiv.org/abs/1902.01603}{{\ttfamily arXiv:1902.01603 [hep-ph]}}.

\bibitem{Furui:2016owe}
A.~Furui, Y.~Hosotani, and N.~Yamatsu, ``{Toward Realistic Gauge-Higgs Grand
  Unification},'' \href{http://dx.doi.org/10.1093/ptep/ptw116}{{ Prog. Theor.
  Exp. Phys.} {\bfseries 2016} (2016) 093B01},
\href{http://arxiv.org/abs/1606.07222}{{\ttfamily arXiv:1606.07222 [hep-ph]}}.

\bibitem{Hosotani:2017ghg}
Y.~Hosotani and N.~Yamatsu, ``{Gauge-Higgs Seesaw Mechanism in 6-Dimensional
  Grand Unification},'' \href{http://dx.doi.org/10.1093/ptep/ptx124}{{ Prog.
  Theor. Exp. Phys.} {\bfseries 2017} no.~9, (2017) 091B01},
\href{http://arxiv.org/abs/1706.03503}{{\ttfamily arXiv:1706.03503 [hep-ph]}}.

\bibitem{Hosotani:2017edv}
Y.~Hosotani and N.~Yamatsu, ``{Electroweak Symmetry Breaking and Mass Spectra
  in Six-Dimensional Gauge-Higgs Grand Unification},''
  \href{http://dx.doi.org/10.1093/ptep/ptx175}{{ Prog. Theor. Exp. Phys.}
  {\bfseries 2018} no.~2, (2018) 023B05},
\href{http://arxiv.org/abs/1710.04811}{{\ttfamily arXiv:1710.04811 [hep-ph]}}.

\bibitem{Englert:2019xhz}
C.~Englert, D.~J. Miller, and D.~D. Smaranda, ``{Phenomenology of GUT-Inspired
  Gauge-Higgs Unification},''
  \href{http://dx.doi.org/10.1016/j.physletb.2020.135261}{{ Phys. Lett. B}
  {\bfseries 802} (2020) 135261},
\href{http://arxiv.org/abs/1911.05527}{{\ttfamily arXiv:1911.05527 [hep-ph]}}.

\bibitem{Englert:2020eep}
C.~Englert, D.~J. Miller, and D.~D. Smaranda, ``{The Weinberg Angle and 5D RGE
  Effects in a SO(11) GUT Theory},''
  \href{http://dx.doi.org/10.1016/j.physletb.2020.135548}{{ Phys. Lett. B}
  {\bfseries 807} (2020) 135548},
  \href{http://arxiv.org/abs/2003.05743}{{\ttfamily arXiv:2003.05743
  [hep-ph]}}.

\bibitem{Funatsu:2013ni}
S.~Funatsu, H.~Hatanaka, Y.~Hosotani, Y.~Orikasa, and T.~Shimotani, ``{Novel
  Universality and Higgs Decay H$\to \gamma\gamma$, gg in the SO(5)$\times$U(1)
  Gauge-Higgs Unification},''
  \href{http://dx.doi.org/10.1016/j.physletb.2013.03.040}{{ Phys. Lett.}
  {\bfseries B722} (2013) 94--99},
\href{http://arxiv.org/abs/1301.1744}{{\ttfamily arXiv:1301.1744 [hep-ph]}}.

\bibitem{Funatsu:2019fry}
S.~Funatsu, H.~Hatanaka, Y.~Hosotani, Y.~Orikasa, and N.~Yamatsu, ``{CKM Matrix
  and FCNC Suppression in $SO(5) \times U(1) \times SU(3)$ Gauge-Higgs
  Unification},'' \href{http://dx.doi.org/10.1103/PhysRevD.101.055016}{{ Phys.
  Rev. D} {\bfseries 101} (2020) 055016},
\href{http://arxiv.org/abs/1909.00190}{{\ttfamily arXiv:1909.00190 [hep-ph]}}.

\bibitem{Funatsu:2020znj}
S.~Funatsu, H.~Hatanaka, Y.~Hosotani, Y.~Orikasa, and N.~Yamatsu, ``{Effective
  Potential and Universality in GUT-Inspired Gauge-Higgs Unification},''
  \href{http://dx.doi.org/10.1103/PhysRevD.102.015005}{{ Phys. Rev. D}
  {\bfseries 102} (2020) 015005},
\href{http://arxiv.org/abs/2002.09262}{{\ttfamily arXiv:2002.09262 [hep-ph]}}.

\bibitem{Funatsu:2020haj}
S.~Funatsu, H.~Hatanaka, Y.~Hosotani, Y.~Orikasa, and N.~Yamatsu, ``{Fermion
  Pair Production at $e^- e^+$ Linear Collider Experiments in GUT Inspired
  Gauge-Higgs Unification},''
  \href{http://dx.doi.org/10.1103/PhysRevD.102.015029}{{ Phys. Rev. D}
  {\bfseries 102} (2020) 015029},
  \href{http://arxiv.org/abs/2006.02157}{{\ttfamily arXiv:2006.02157
  [hep-ph]}}.

\bibitem{Funatsu:2021gnh}
S.~Funatsu, H.~Hatanaka, Y.~Hosotani, Y.~Orikasa, and N.~Yamatsu,
  ``{Electroweak and Left-Right Phase Transitions in $SO(5) \times U(1) \times
  SU(3)$ Gauge-Higgs Unification},''
  \href{http://dx.doi.org/10.1103/PhysRevD.104.115018}{{ Phys. Rev. D}
  {\bfseries 104} (4, 2021) 115018},
  \href{http://arxiv.org/abs/2104.02870}{{\ttfamily arXiv:2104.02870
  [hep-ph]}}.

\bibitem{Funatsu:2021yrh}
S.~Funatsu, H.~Hatanaka, Y.~Hosotani, Y.~Orikasa, and N.~Yamatsu, ``{Signals of
  $W'$ and $Z'$ Bosons at the LHC in the $SU(3) \times SO(5) \times U(1)$
  Gauge-Higgs Unification},''
  \href{http://dx.doi.org/10.1103/PhysRevD.105.055015}{{ Phys. Rev. D}
  {\bfseries 105} (2022) 055015},
  \href{http://arxiv.org/abs/2111.05624}{{\ttfamily arXiv:2111.05624
  [hep-ph]}}.

\bibitem{Funatsu:2022spb}
S.~Funatsu, H.~Hatanaka, Y.~Hosotani, Y.~Orikasa, and N.~Yamatsu, ``{Bhabha
  Scattering in the Gauge-Higgs Unification},'' { Phys. Rev. D} {\bfseries 106}
  (3, 2022) 015015, \href{http://arxiv.org/abs/2203.16030}{{\ttfamily
  arXiv:2203.16030 [hep-ph]}}.

\bibitem{Funatsu:2023jng}
S.~Funatsu, H.~Hatanaka, Y.~Orikasa, and N.~Yamatsu, ``{Single Higgs Boson
  Production at Electron-Positron Colliders in Gauge-Higgs Unification},''
  \href{http://dx.doi.org/10.1103/PhysRevD.107.075030}{{ Phys. Rev. D}
  {\bfseries 107} (1, 2023) 075030},
  \href{http://arxiv.org/abs/2301.07833}{{\ttfamily arXiv:2301.07833
  [hep-ph]}}.

\bibitem{Hosotani:2023yhs}
Y.~Hosotani, S.~Funatsu, H.~Hatanaka, Y.~Orikasa, and N.~Yamatsu, ``{Coupling
  Sum Rules and Oblique Corrections in Gauge-Higgs Unification},''
  \href{http://dx.doi.org/10.1093/ptep/ptad068}{{ PTEP} {\bfseries 2023} no.~6,
  (2023) 063B01}, \href{http://arxiv.org/abs/2303.16418}{{\ttfamily
  arXiv:2303.16418 [hep-ph]}}.

\bibitem{Yamatsu:2023bde}
N.~Yamatsu, S.~Funatsu, H.~Hatanaka, Y.~Hosotani, and Y.~Orikasa, ``{$W$ and
  $Z$ Boson Pair Production at Electron-Positron Colliders in Gauge-Higgs
  Unification},'' { Phys. Rev. D} {\bfseries 108} (9, 2023) 115014,
  \href{http://arxiv.org/abs/2309.01132}{{\ttfamily arXiv:2309.01132
  [hep-ph]}}.

\bibitem{Hosotani:2023poh}
Y.~Hosotani, S.~Funatsu, H.~Hatanaka, Y.~Orikasa, and N.~Yamatsu, ``{W Boson
  Mass in Gauge-Higgs Unification},''
  \href{http://dx.doi.org/10.1103/PhysRevD.108.115036}{{ Phys. Rev. D}
  {\bfseries 108} no.~11, (2023) 115036},
  \href{http://arxiv.org/abs/2310.03276}{{\ttfamily arXiv:2310.03276
  [hep-ph]}}.

\bibitem{Irles:2024ipg}
A.~Irles, J.~P. M\'arquez, R.~P\"oschl, F.~Richard, A.~Saibel, H.~Yamamoto, and
  N.~Yamatsu, ``{Probing Gauge-Higgs Unification Models at the ILC with
  Quark\textendash{}Antiquark Forward\textendash{}Backward Asymmetry at
  Center-of-Mass Energies above the Z mass},''
  \href{http://dx.doi.org/10.1140/epjc/s10052-024-12918-z}{{ Eur. Phys. J. C}
  {\bfseries 84} no.~5, (2024) 537},
  \href{http://arxiv.org/abs/2403.09144}{{\ttfamily arXiv:2403.09144
  [hep-ph]}}.

\bibitem{Koide:1987vp}
Y.~Koide and S.~Oneda, ``{Lepton Masses and SU(3) Family Symmetry Breaking},''
  \href{http://dx.doi.org/10.1103/PhysRevD.36.2867}{{ Phys. Rev. D} {\bfseries
  36} (1987) 2867}.

\bibitem{Koide:1995xk}
Y.~Koide and M.~Tanimoto, ``{U(3) Family Nonet Higgs Boson and its
  Phenomenology},'' \href{http://dx.doi.org/10.1007/BF02909162}{{ Z. Phys. C}
  {\bfseries 72} (1996) 333--344},
  \href{http://arxiv.org/abs/hep-ph/9505333}{{\ttfamily arXiv:hep-ph/9505333}}.

\bibitem{Koide:2008ey}
Y.~Koide, ``{U(3)-Flavor Nonet Scalar as an Origin of the Flavor Mass
  Spectra},'' \href{http://dx.doi.org/10.1016/j.physletb.2008.02.059}{{ Phys.
  Lett. B} {\bfseries 662} (2008) 43--48},
  \href{http://arxiv.org/abs/0802.1084}{{\ttfamily arXiv:0802.1084 [hep-ph]}}.

\bibitem{Koide:2013eca}
Y.~Koide and H.~Nishiura, ``{Yukawaon Model with Anomaly Free Set of Quarks and
  Leptons in a U(3) Family Symmetry},''
  \href{http://dx.doi.org/10.1103/PhysRevD.88.116004}{{ Phys. Rev. D}
  {\bfseries 88} no.~11, (2013) 116004},
  \href{http://arxiv.org/abs/1308.2129}{{\ttfamily arXiv:1308.2129 [hep-ph]}}.

\bibitem{Koide:2014doa}
Y.~Koide, ``{Spectroscopy of Family Gauge Bosons},''
  \href{http://dx.doi.org/10.1016/j.physletb.2014.07.061}{{ Phys. Lett. B}
  {\bfseries 736} (2014) 499--505},
  \href{http://arxiv.org/abs/1405.6778}{{\ttfamily arXiv:1405.6778 [hep-ph]}}.

\bibitem{Koide:2019bgk}
Y.~Koide, ``{SU(5)$_{\rm flavor}$ \texttimes{}SU(3)$_{\rm family}$ Model with
  Unconventional Family Assignment},''
  \href{http://dx.doi.org/10.1016/j.physletb.2019.134909}{{ Phys. Lett. B}
  {\bfseries 797} (2019) 134909}.

\bibitem{Ross:2002fb}
G.~G. Ross and L.~Velasco-Sevilla, ``{Symmetries and Fermion Masses},''
  \href{http://dx.doi.org/10.1016/S0550-3213(03)00041-5}{{ Nucl. Phys. B}
  {\bfseries 653} (2003) 3--26},
  \href{http://arxiv.org/abs/hep-ph/0208218}{{\ttfamily arXiv:hep-ph/0208218}}.

\bibitem{King:2003rf}
S.~F. King and G.~G. Ross, ``{Fermion Masses and Mixing Angles from $SU(3)$
  Family Symmetry and Unification},''
  \href{http://dx.doi.org/10.1016/j.physletb.2003.09.027}{{ Phys. Lett.}
  {\bfseries B574} (2003) 239--252},
\href{http://arxiv.org/abs/hep-ph/0307190}{{\ttfamily arXiv:hep-ph/0307190}}.

\bibitem{Appelquist:2006ag}
T.~Appelquist, Y.~Bai, and M.~Piai, ``{Quark Mass Ratios and Mixing Angles from
  SU(3) Family Gauge Symmetry},''
  \href{http://dx.doi.org/10.1016/j.physletb.2006.04.043}{{ Phys. Lett. B}
  {\bfseries 637} (2006) 245--250},
  \href{http://arxiv.org/abs/hep-ph/0603104}{{\ttfamily arXiv:hep-ph/0603104}}.

\bibitem{Appelquist:2006qq}
T.~Appelquist, Y.~Bai, and M.~Piai, ``{Neutrinos and SU(3) family gauge
  symmetry},'' \href{http://dx.doi.org/10.1103/PhysRevD.74.076001}{{ Phys. Rev.
  D} {\bfseries 74} (2006) 076001},
  \href{http://arxiv.org/abs/hep-ph/0607174}{{\ttfamily arXiv:hep-ph/0607174}}.

\bibitem{Yang:2011qj}
W.-M. Yang, Q.~Wang, and J.-J. Zhong, ``{A Model of Fermion Masses and Flavor
  Mixings with Family Symmetry $SU(3)\otimes U(1)$},''
  \href{http://dx.doi.org/10.1088/0253-6102/57/1/12}{{ Commun. Theor. Phys.}
  {\bfseries 57} (2012) 71--77},
  \href{http://arxiv.org/abs/1108.4512}{{\ttfamily arXiv:1108.4512 [hep-ph]}}.

\bibitem{Wu:2012ria}
Y.-L. Wu, ``{SU(3) Gauge Family Symmetry and Prediction for the Lepton-Flavor
  Mixing and Neutrino Masses with Maximal Spontaneous CP Violation},''
  \href{http://dx.doi.org/10.1016/j.physletb.2012.07.020}{{ Phys. Lett. B}
  {\bfseries 714} (2012) 286--294},
  \href{http://arxiv.org/abs/1203.2382}{{\ttfamily arXiv:1203.2382 [hep-ph]}}.

\end{thebibliography}\endgroup

\end{document}